\documentclass[12pt,epsfig]{article}
\usepackage{graphicx}
\usepackage{amsmath}
\usepackage{amsfonts}
\usepackage{amssymb}
\def\beq{\begin{equation}}
\def\eeq{\end{equation}}
\def\bea{\begin{eqnarray}}
\def\eea{\end{eqnarray}}

\begin{document}

\begin{center}
{\Large \bf New scattering features of quaternionic point interaction in non-Hermitian quantum mechanics 
  }

\vspace{1.3cm}

{\sf   Mohammad Hasan \footnote{e-mail address: \ \ mohammadhasan786@gmail.com, \ \ mhasan@isro.gov.in}
and Bhabani Prasad Mandal \footnote{e-mail address:
\ \ bhabani.mandal@gmail.com, \ \ bhabani@bhu.ac.in  }}

\bigskip

{\em $^1$ISRO Head Quarters ,
Bangalore-560231, INDIA \\
$^{1, 2}$Department of Physics,
Banaras Hindu University,
Varanasi-221005, INDIA. \\ }

\bigskip
\bigskip

\noindent {\bf Abstract}
 
\end{center}
The spectral singularity have been extensively studied over the last one and half decade for different non-Hermitian potentials in non-Hermitian quantum mechanics. The nature of spectral singularities have not been studied for the case of quaternionic potential. In the present work we perform an analytical study for the scattering from a quaternionic point interaction represented by delta function. New features of spectral singularities are observed  which are different than the case of complex (non-quaternionic) point interaction. Most notable difference is the occurrence of spectral singularity from lossy point interaction which is forbidden in the case of standard non-Hermitian quantum mechanics.

\medskip
\vspace{1in}
\newpage

\section{Introduction}
The early work of Birkhoff and Neumann \cite{birkhoff_neumann_paper_1936, neumann_book} led to the possibility of formulating the quantum mechanics over quaternions. Since then it remains an open question whether quantum mechanical dynamics require a generalization of complex numbers over quaternion numbers or over other hyper-complex numbers. Does nature prefers complex numbers over quaternions is an experimental problem to be probed. However to enable this an accurate prediction based on theoretical grounds are required to understand the experimental domain where a departure in predicted outcome of Hermitian quantum mechanics (HQM) and quaternionic quantum mechanics (QQM) is expected in experimental results. Various authors have made their contributions towards this.
\paragraph{}
The generalization of HQM to QQM doesn't imply additional degree of freedom in the theory. QQM  is fundamentally different than HQM but does not modify the postulate of quantum mechanics. One of the disagreement between HQM and QQM is the commutativity of phases. Quaternions rotation does not commute, thus the phases in QQM shows non-commutativity. In 1979, Asher Peres proposed several experiments (Bragg scattering, neutron diffraction and $K_{s}$ meson regenerations) \cite {asher_peres_1979} for testing the presence of quaternionic effects by observing non-commutativity of phases. The non-commutativity of phases at the wave function level was shown by Davies et al.  \cite{davies_1992}.  In 1984 , Kaiser et al. performed the first experiment using neutron interferometry to search for quaternions in quantum mechanics \cite{kaiser_1984}. In this experiment  a phase change was looked in the transmitting beams of neutron by reversing the order of two metal targets. No such phase difference  was found. It was later shown that the quaternionic effects decay exponentially for massive particles \cite {adler_book} and thus the neutron interferometry experiments did not probe the quaternionic effects.  Adler pointed towards the CP-Nonconservation interaction at mass scale of about 20 TeV as a result of underlying quaternionic effects in particle interaction \cite{adler_superweak_cp}. Due to technological limitations no further test were performed until 2016. Procopio et al. \cite{procopio_2016} performed single photon test by using a Sagnac interferometer. The two very different optical media used were liquid-crystal and fishnet metamaterial with negative refractive index (at 780 nm). The phase shift between the two modes of the Sagnac interferometer yield a bound of $\theta=0.03^{\circ}$ indicating that phase commutes with high accuracy. See \cite{adler_2017, procopio_2017} for critical comments on the experiment.
\paragraph{}
For the case when quaternion Hamiltonian is anti-Hermitian, the probability is conserved and unitarity condition is satisfied in the scattering problems. This makes the QQM as the theory of anti-Hermitian operators \cite{adler_book}. The scattering problems of anti-Hermitian scalar potential in QQM have been studied for relativistic and non-relativistic cases both. Due to the mathematical complexity involved with quaternionic algebra, these problems have been limited for simple potential configurations  such as delta \cite{d1,d2}, double delta \cite{d2_1}, rectangular \cite {b1, b2} and step potential \cite {s1,s2}. To the best of our knowledge the scattering properties from a non-anti-Hermitian quaternionic quantum potential has not been studied so far. 
\paragraph{}
The motivation for the present work arise due to the recent works by S. Giardino \cite{giardino_abe, giardino_nah} and the tight (small) experimental bound on the non-commutativity of phases in QQM \cite{procopio_2016}.  Ref \cite{giardino_abe} posed the question on the necessity of anti-Hermitian assumption in QQM. Ref \cite{giardino_nah} propose a non-anti-Hermitian QQM as quaternionic generalization of previously known (non-quaternionic) non-Hermitian quantum mechanics (NHQM). The non-Hermitian systems with real energy eigenvalues has become the topic of frontier research over the last two decades. It is because that one can have a consistent quantum theory by restoring the Hermiticity and upholding the unitary time evolution for such a system in a modified Hilbert space \cite{ben4}-\cite{benr}. The non-Hermitian Hamiltonian (NHQM) shows several new features which are absent in the case of Hermitian Hamiltonian. These are exceptional points (EPs) \cite{ep0}-\cite{ep2}, spectral singularity (SS) \cite{ss1}-\cite{ss3}, invisibility  \cite{aop}-\cite{inv3}, reciprocity  \cite{aop}-\cite{resc}, coherent perfect absorption (CPA) \cite{cpa00}-\cite{cpa4}, critical coupling (CC) \cite{cc0}-\cite{cc4} and CPA-laser \cite{cpa_laser1}.   In optical domain, some of the features have been verified experimentally \cite{ opt1}-\cite{eqv1}. The experimental realization of the predictions of non-Hermitian theory have further boosted huge interest in the study of non-Hermitian system in experimental as well as in theoretical domains. Recently CPA and SS were also studied in the context of non-Hermitian fractional quantum mechanics \cite{nhsfqm}. 
\paragraph{}
The experimental result of \cite{procopio_2016} strongly indicate the need for the exploration of more sensitive deviation of quaternionic effects. One such domain to search for sensitive deviation is the study of spectral singularity in the scattering problem of non-anti-Hermitian QQM.  Spectral singularities are characterize by the simultaneous divergence of reflection and transmission amplitude from a non-Hermitian potential (see \cite{ali_m_chapter} for detail discussion and references therein). These are also called as zero width resonance and found to be extremely sensitive to the dimension of interacting region \cite {ali_ss_barrier_2009}. As stated above it is  proposed that \cite{giardino_nah} non-anti-Hermitian QQM can be possibly formulated as a generalization of NHQM, it is most desirable to study the scattering features in the domain of non-anti-Hermitian QQM and study for the possible deviation in the scattering features between non-anti-Hermitian QQM and NHQM. To provide analytical understanding (and thus avoiding transcendental equations), we chose point interaction represented by quaternionic (single) delta potential for our study of spectral singularity in non-anti-Hermitian QQM. The objective of the present work is not to propose an experiment to test the quaternionic effects rather the analytical understanding of the deviation of scattering features (SS here) between non-anti-Hermitian QQM and NHQM. We found the new features of SS occurrence in non-anti-Hermitian QQM which are forbidden in the case of standard NHQM. 
\paragraph{}
We organize the paper as follows: In section \ref{qqm_intro} we briefly introduce the reader with quaternionic quantum mechanics. Section \ref{scattering_delta_qqm} present our detail analytical calculation for the investigation of spectral singularity from quaternionic delta potential. In this section we also discuss new features of SS and the limiting cases. The results are discussed in section \ref{resultt}. Throughout the test we use `non-Hermitian QQM' to distinguish it from `anti-Hermitian QQM'. 

\section{Quaternionic quantum mechanics}
\label{qqm_intro}
The non-relativistic quaternionic quantum mechanics is governed by quaternionic form of Schrodinger equation. The quaternionic Schrodinger equation is 

\begin{equation}
i \hbar \frac{\partial \psi (x,t)}{\partial t}= - \frac{\hbar^2}{2m} \nabla^{2} \psi(x,t) +V(x,t) \psi(x,t)
\label{tdqse}
\end{equation}
where $V(x,t)$ is non-Hermitian scalar potential and is expressed in terms of quaternion. $V(x,t)$ is given by
\beq
V(x,t)=V_{0}+iV_{1}(x,t)+jV_{2}(x,t)+kV_{3}(x,t)
\eeq
Here $\{V_{0}(x,t), V_{1}(x,t),V_{2}(x,t),V_{3}(x,t)\} \in R$ and $i,j,k$ are three imaginary units with the properties $i^2=j^2=k^2=-1$. Also $ijk=-1$. The property $ijk=-1$ makes quaternion numbers non commutative.

When potential $V(x,t)=V(x)$ i.e. independent of time we have the time independent quaternionic Schrodinger equation. This is given by 
\begin{equation}
- \frac{\hbar^2}{2m} \nabla^{2} \psi(x) +V(x) \psi(x)=0
\label{qse}
\end{equation}
where now $V(x)=V_{0}(x)+iV_{1}(x)+jV_{2}(x)+kV_{3}(x)$ and $\{V_{0}(x), V_{1}(x),V_{2}(x),V_{3}(x)\} \in R$. The imaginary units $i,j,k$ have the same properties as defined above. It is evident that in the limit $V_{1} \rightarrow 0$, $V_{2} \rightarrow 0$, $V_{3} \rightarrow 0$ we  recover the standard Schroedinger equation.

\section{Scattering features from non-Hermitian delta potential in quaternionic quantum mechanics} 
\label{scattering_delta_qqm}
The scattering coefficient from an anti-Hermitian quaternionic Dirac delta potential in one dimension
\beq
V(x)=(iV_{1}+jV_{2}+kV_{3})\delta (x)
\label{anti_delta}
\eeq
 $\{V_{1},V_{2},V_{3}\} \in R$ has been computed in \cite{d1}. The reflection and transmission coefficients are given by
\begin{equation}
r=- \frac{i}{D}\left( V_{1}^{2}+g^{2}+ V_{1} \beta \right)
\label{r1}
\end{equation}
and
\beq
t=\frac{1}{D} \beta ( \beta +V_{1})
\label{t1}
\eeq 
where
\beq
D=\beta(\beta+V_{1})+i(V_{1}^{2}+g^{2}+ V_{1} \beta )
\label{denomr}
\eeq

$\beta=\frac{\hbar^{2} \epsilon}{m}$, $\epsilon= \frac{\sqrt{2mE}}{\hbar}$ and $g^{2}=V_{2}^{2}+V_{3}^{2}>0$ for $V_{2},V_{3} \in R$.
When $V_{1}\in C$ then the quaternionic delta potential given by eq. \ref{anti_delta} is non-anti-Hermitian delta potential. For simplicity we consider $V_{2},V_{3} \in R$. Thus our non-anti-Hermitian delta potential is 
\beq
V(x)=(iV_{1}+jV_{2}+kV_{3})\delta (x)
\label{non_anti_delta}
\eeq
with $V_{1}\in C$ and $V_{2},V_{3} \in R$. From eq. \ref{r1} and \ref{t1} we see that for spectral singularity we must have $D=0=\vert D \vert^{2} $. Substituting $V=v_{1}+iv_{2}$, $v_{1},v_{2} \in R$ in eq. \ref{denomr} we obtain
\begin{equation}
\vert D\vert^{2}= a\beta^{4}+b\beta^{3}+c\beta^{2}+d\beta+e
\end{equation}  
Here 

\begin{equation}
a=1
\label{a_value}
\end{equation}
\begin{equation}
b=2(v_{1}-v_{2})
\label{b_value}
\end{equation}
\begin{equation}
c=2 (v_{1}-v_{2})^{2}
\label{c_value}
\end{equation}
\begin{equation}
d=2 \{g^2 ({v_{1}}+{v_{2}})+({v_{1}}-{v_{2}}) ({v_{1}}^2+{v_{2}}^2)\}
\label{d_value}
\end{equation}
\begin{equation}
e=g^4+2 g^2 ({v_{1}}^2-{v_{2}}^2)+({v_{1}}^2+{v_{2}}^2)^2
\label{e_value}
\end{equation}
Therefore to find the spectral singularity we have to solve the following quartic equation
\beq
\beta^{4}+b\beta^{3}+c\beta^{2}+d\beta+e=0
\label{quartic_eq}
\eeq
for $\beta$. As we have to find the positive energy at which spectral singularity occurs we have to solve the above equation for roots such that $\beta \in R^{+}$. To find the nature of the root we first investigate the determinant of eq. \ref{quartic_eq}. The determinant of quartic eq. \ref{quartic_eq} is well known and is given by 

\begin{multline}
\Delta = 256 e^{3}-192 bde^{2} -128c^{2}e^{2} +144cd^{2}e- 27d^{4}+ 144b^{2}ce^{2}- 6b^{2}d^{2}e- 80bc^{2}de+ \\  18bcd^{3} + 16c^{4}e - 4c^{3}d^{2} - 27b^{4}e^{2} + 18b^{3}cde - 4b^{3}d^{3}  - 4b^{2}c^{3}e + b^{2}c^{2}d^{2}      
\label{discriminant}
\end{multline}

Substituting the values of $b,d,e,f$ in the expression of determinant we have for the current problem
\beq
\Delta = 64.A.B
\label{determiant_simplified}
\eeq
where
\beq
A=[g^4+g^2 (v_{1}^2-v_{2}^2) -2 {v_{1}} {v_{2}} ({v_{1}}+{v_{2}})^2]^2
\label{a_expression} 
\eeq 
\beq
B=4 g^4+4 g^2 (v_{1}^2-v_{2}^2) +(v_{1}^2+v_{2}^2)^2
\label{b_expression}
\eeq 
From the properties of quartic function it is known that if $\Delta<0$ then two roots are distinctly real. For the case $\Delta >0$, then either all the four roots are real or none is. In this case the nature of the roots depend upon the sign of 
\begin{equation}
P=8c-3b^{2}
\label{p_eq}
\end{equation}
and 
\begin{equation}
Q=64e-16c^{2}+16b^{2}c-16bd-3b^{4}
\label{d_eq}
\end{equation}
If $P<0$ and $Q<0$, then all the roots are real. From eq. \ref{a_expression} we see that $A \geq 0$. Thus the sign of the determinant is decided by the sign of $B$.
Substituting the values of $b,c,d,e$ in eq. \ref{p_eq} and \ref{d_eq} we have
\begin{equation}
P=4 (v_{1}-v_{2})^{2}
\label{p_value}
\end{equation}
and
\begin{equation}
Q=16 [4 g^4+4 g^2(v_{1}^{2}-v_{2}^{2}) +(v_{1}+v_{2})^4]
\label{q_value}
\end{equation}
From eq. \ref{p_value} we observe $P \geq 0$. Thus if $\Delta>0$ or $B>0$ there is no real number $\beta$ which satisfies  eq. \ref{quartic_eq}.
Now we investigate the possibility of $\Delta <0$. For this we must have $B<0$ i.e.
\beq
4 g^4+4 g^2 (u-v) +(u+v)^2 <0
\label{uv_inequality}
\eeq
with $u=v_{1}^2$ and $v=v_{2}^2$.
For $g,v \in R^{+}$ the inequality \ref{uv_inequality} is satisfied for the range of $u$ given by
\beq
-2g^2-v-2g\sqrt{2v} <u< -2g^2-v+2g\sqrt{2v} 
\eeq
which is equivalent to
\beq
- (\sqrt{2}g+v)^2 <u < -(\sqrt{2}g-v)^2 
\eeq
This shows that there doesn't exist $u>0$ for which inequality \ref{uv_inequality} can be satisfied. This essentially indicates that for a chosen pair $v_{1},v_{2} \in R$, such that $V_{1}=v_{1}+iv_{2}$, the quaternionic delta potential doesn't show spectral singularity. However it must be noted that the above arguments is not entirely accurate in ruling out the existence of SS in quaternionic delta potential. The logics and calculated results presented above shows that for a `{\it free}' set of chosen numbers $\{v_{1},v_{2}, g^{2}=V_{2}^{2}+V_{3}^{2} \} \in R$, one may not get SS in quaternionic delta potential. To investigate further we write
\beq
D=D_{r}+i D_{i}
\eeq
where $\{D_{r}, D_{i} \} \in R$  

\begin{figure}
\begin{center}
\includegraphics[scale=0.7]{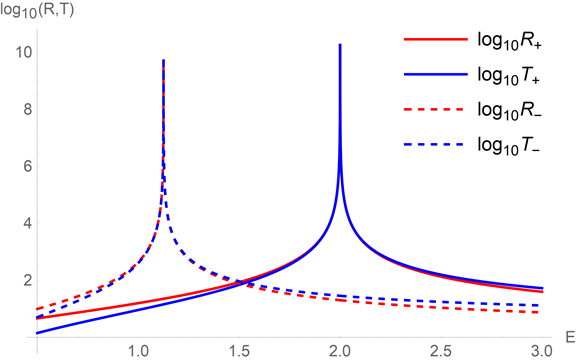} 
\caption{\it Graphical representation of spectral singularity for the quaternionic delta potential $V=-10-0.5i+j V_{2}+k V_{3}$ for the two configuration of quaternionic part , where $g_{+}^{2}=V_{2}^{2}+ V_{3}^{2}= \frac{15}{4}$ and $g_{-}^{2}=V_{2}^{2}+ V_{3}^{2}= 5$ as calculated from eq. \ref{g_exp}. $ \{ V_{2}, V_{3} \} \in R $. The corresponding spectral singularity occur at  $E_{+}^{ss}=2 $ and $E_{-}^{ss}=\frac{9}{8} $ as calculated from eq. \ref{ess}. $R_{\pm}, T_{\pm}$ correspond to the case when quaternionic part are given by $g_{\pm}^{2}$.  Solid lines corresponds to the case of $g_{+}^{2}$ while dotted lines corresponds to the case of $g_{-}^{2}$. } 
\label{ss_v1_lessthan_zero}
\end{center}
\end{figure}

For SS to  occur $D_{r}$, $D_{i}$ must vanish simultaneously. Thus $\beta \in R^{+}$ must satisfy
\beq
\beta^{2}+\beta (v_{1}-v_{2}) -2 v_{1}v_{2}=0 
\label{dr}
\eeq  
and,
\beq
v_{1}^{2}-v_{2}^{2} +\beta (v_{1}+v_{2}) +g^{2}=0 
\label{di}
\eeq  
Eq. \ref{dr} is independent of $g$ while eq. \ref{di} is dependent on $g$. This shows that for $\beta$ to satisfy above two equations, $g$ can not be a `free' parameter and must depend on $v_{1}, v_{2}$. Solutions of eqs. \ref{dr} and \ref{di} gives allowed value of $g$ as
\beq
g_{\pm}^{2}= -\frac{1}{2}(v_{1}+v_{2})\left((v_{1}-v_{2}) \pm \sqrt{(v_{1}+v_{2})^{2} +4 v_{1}v_{2}}\right)
\label{g_exp}
\eeq   
The quaternionic delta potential depict SS at energy.   
\beq
E^{ss}_{\pm}=\frac{1}{2} \left ((v_{2}-v_{1}) -\frac{g_{\pm}^{2}}{(v_{1}+v_{2})} \right)^{2}
\label{ess}
\eeq 
where $g_{\pm}$ is given from eq. \ref{g_exp}. In the above we have used $\hbar=1, m=1$. It is to be noted that for a given pair $\{v_{1}, v_{2}\}$, not all values of $g_{\pm}^{2}=V_{2}^{2}+V_{3}^{2}$ obtained from eq. \ref{g_exp} will support SS. As $V_{2}, V_{3} \in R$, this implies that $g_{\pm}^{2}>0$. The other restriction is $\beta_{\pm}= \sqrt{2 E_{ss}^{\pm}} >0$. The expressions for $\beta_{\pm}$ is given below which can be obtained by substituting $g_{\pm}^{2}$ in eq. \ref{di}. 
\beq
\beta_{\pm}= -(v_{1}-v_{2}) -  \frac{g_{\pm}^{2}}{v_{1}+v_{2}} 
\label{beta_pm}
\eeq 
With the restrictions mentioned above, we investigate the parameter space  of $\{ v_{1}, v_{2} \}$ for which the quaternionic delta system support SS. Imposing the condition $g_{+}^{2}>0$ and $\beta_{+} > 0$, we see that for $v_{2}<0$, the allowed values of $v_{1}$ are $v_{1} <0$. For $v_{2}>0$ the range of $v_{1}$ is $(-3+2 \sqrt{2}) v_{2}\leq v_{1}<0$ to support SS. For condition $g_{-}^{2}>0$ and $\beta_{-} > 0$ to satisfy simultaneously the only parameter space are $(-3+2 \sqrt{2}) v_{2}\leq v_{1}<0$ where $v_{2}>0$. The most interesting outcome is to note that SS exists even when complex part $v_{1}<0$ (as $i V_{1}=-v_{2}+i v_{1}$). This is forbidden in standard non-Hermitian quantum mechanics. Further for a given $g_{\pm}^{2}>0$ one can have a lossy point interaction as $g^{2}= (- V_{2})^{2}+ (- V_{3})^{2}=(+ V_{2})^{2}+ (+ V_{3})^{2}$. Thus a suitably chosen lossy quaternionic delta potential can exhibit SS. In general a lossy quaternionic delta potential $V(x)= (v_{2}-i v_{1} -j V_{2}-k V_{3}) \delta(x)$ with $\{ v_{1}, v_{2} ,V_{2}, V_{3} \} \in R^{+}$ always shows SS at energy given from eq. \ref{ess} : at $E_{+}^{ss}$ provided $g_{+}^{2}$ is given from eq. \ref{g_exp}. The graphical representation of SS for $v_{1}<0$ is shown in Fig \ref{ss_v1_lessthan_zero} for the case when quaternionic part are given from $g_{\pm}^{2} >0$ with the condition $\beta_{\pm} >0$. In  both cases ($g_{+}$ and $g_{-}$), SS occur at energy calculated from eq. \ref{ess} which verifies our calculations. It is also observed that SS can also exists even when real part of quaternionic delta potential is greater than zero which is forbidden in case of non-quaternionic complex delta potential \cite{ss_d}. 
It can be verified that for $v_{1} \rightarrow 0^{-}$, we have $g_{+}^{2} \rightarrow 0$ and thus the quaternionic delta potential reduces to a pure complex delta potential $\delta(x)=-i v_{2}$. As discussed above $v_{1} \rightarrow 0^{-}$, (i.e. $V_{1}<0$) will have SS provided $v_{2} <0$. This means that imaginary part of $\delta(x)$ is positive. Further using eq. \ref{ess} it can be shown that SS occur at energy $E_{+}^{ss}=\frac{v_{2}^{2}}{2}$ which is the well known result of standard NHQM \cite{ss_d}. This verifies our calculation in appropriate limit. For the other solution ($g_{-}^{2}$), $v_{1} \rightarrow 0^{-}$ implies $g_{-}^{2} \rightarrow v_{2}^{2}$ and correspondingly $E_{-}^{ss} \rightarrow \frac{1}{2} v_{1}^{2}$. In this limiting case $E_{-}^{ss}$ is largely independent of $v_{2}$ and thus independent on $g_{-}^{2}$. Thus a quaternion delta potential having only $j$ and $k$ components (along with real component) doesn't support SS ($E_{-}^{ss}=0$ in this case). Further a finite quaternion delta potential $V(x)= v_{2}-i v_{1} -j V_{2}-k V_{3}$ such that $v_{1} \rightarrow 0^{-}$ and $V_{2}^{2}+ V_{3}^{2} \sim v_{2}^{2}$ support SS at a near-zero energy given by $E_{-}^{ss}= 0.5 v_{1}^{2}$. This is different than the results of standard NHQM in which case a near-zero energy SS also means vanishing strength of imaginary delta potential \cite{ss_d}. 

\section{Conclusions and Discussions}
\label{resultt}
In this work we have analytically studied the occurrence of spectral singularity (SS) from a point interaction represented by quaternionic delta potential. It is shown that the quaternionic delta potential support SS for specific values of the parameters of the potential. Unlike the case of standard NHQM, the quaternion delta potential can support SS even when $Re[\delta (x)])\neq 0$ where $Re$ represent the real part. It is also observed that SS occur when $Im^{i}[\delta (x)]<0$ where $Im^{i}$ represent the coefficient of imaginary number $i$. This case is not supported in standard NHQM for the occurrence of SS as it represent a lossy interaction.  It is shown further that our results of non-Hermitian QQM correctly reduces to the limiting case of standard NHQM. A delta interaction with large quaternionic strength can support SS at vanishingly small energy in non-Hermtian QQM. This is again different than the delta interaction of standard NHQM in which occurrence of SS at vanishingly small energy demands vanishing small strength of delta interaction as well \cite{ss_d}.  We hope that understanding the deviations of non-Hermitian QQM to that of standard NHQM will pave way for new methodologies to search for quaternionic effects in quantum mechanics.

{\it \bf{Acknowledgements}}: \\
MH acknowledges support from Director, SSPO/ ISRO  to carry out this research work. BPM acknowledges the support from CAS, Department of Physics, BHU.

\end{document}